\documentclass[eqsecnum,aps,showpacs]{revtex4}

\begin{document}

\begin{flushright}
DOE/ER/40762-276\\
UMPP\#03-037
\end{flushright}

\count255=\time\divide\count255 by 60 \xdef\hourmin{\number\count255}
  \multiply\count255 by-60\advance\count255 by\time
 \xdef\hourmin{\hourmin:\ifnum\count255<10 0\fi\the\count255}

\newcommand{\xbf}[1]{\mbox{\boldmath $ #1 $}}

\newcommand{\sixj}[6]{\mbox{$\left\{ \begin{array}{ccc} {#1} & {#2} &
{#3} \\ {#4} & {#5} & {#6} \end{array} \right\}$}}

\newcommand{\threej}[6]{\mbox{$\left( \begin{array}{ccc} {#1} & {#2} &
{#3} \\ {#4} & {#5} & {#6} \end{array} \right)$}}

\title{Excited Baryons in Large $N_c$ QCD Revisited: The Resonance Picture
Versus Single-Quark Excitations}

\author{Thomas D. Cohen}
\email{cohen@physics.umd.edu}

\affiliation{Department of Physics, University of Maryland, College
Park, MD 20742-4111}

\author{Richard F. Lebed}
\email{Richard.Lebed@asu.edu}

\affiliation{Department of Physics and Astronomy, Arizona State
University, Tempe, AZ 85287-1504}

\date{January, 2003}

\bigskip

\begin{abstract}
We analyze excited baryon properties via a $1/N_c$ expansion from two
perspectives: as resonances in meson-nucleon scattering, and as
single-quark excitations in the context of a simple quark model.  For
both types of analysis one can derive novel patterns of degeneracy
that emerge as $N_c \rightarrow \infty$, and that are shown to be
compatible with one another.  This helps justify the single-quark
excitation picture and may give some insight into its successes.  We
also find that in the large $N_c$ limit one of the $S_{11}$ baryons
does not couple to the $\pi$-$N$ channel but couples to the $\eta$-$N$
channel.  This is empirically observed in the $N(1535)$, which couples
very weakly to the $\pi$-$N$ channel and quite strongly to the
$\eta$-$N$ channel.  The comparatively strong coupling of the
$N(1650)$ to the $\pi$-$N$ channel and weak coupling to $\eta$-$N$
channel is also predicted.  In the context of the simple quark model
picture we reproduce expressions for mixing angles that are accurate
up to $O(1/N_c)$ corrections and are in good agreement with mixing
angles extracted phenomenologically.
\end{abstract}

\pacs{11.15.Pg, 12.39.-x, 13.75.Gx, 14.20.Gk}

\maketitle
\thispagestyle{empty}

\newpage
\setcounter{page}{1}

\section{Introduction}\label{sec:intro}

The intellectual history of the quark model is a study in irony.  In
the 1960's the quark model provided a concrete dynamical model that
incorporated SU(3) flavor in a natural way, and thus organized the
known hadrons according to an intelligible scheme.  As such, it
provided an essential role in the evolution of ideas that ultimately
led to the formulation of QCD in the early 1970's as the underlying
theory of strong-interaction dynamics.  However, once QCD
was discovered, the status of the quark model became quite
problematic, since it was not known how this model was connected to
QCD, as there was no known way to derive the quark model from QCD as
some type of approximate treatment of the full theory.  This irony
would be of interest only to historians of science, if not for the
fact that fully three decades since the QCD Lagrangian was first
written down the quark model remains the principal tool used by the
community for describing hadronic resonances.  The reason is simple:
Direct {\it ab initio\/} calculation of excited states from QCD
remains an exceptionally difficult problem (although new lattice
techniques are showing some promise~\cite{lattice}).

This paper focuses on the relationship of the quark model treatment of
excited baryons to QCD, using techniques based on the $1/N_c$
expansion~\cite{Hoo,Wit}.  We study excited baryons because the
contracted SU($2N_f$) spin-flavor symmetry that emerges for baryons in
large $N_c$~\cite{GS,DM} provides a powerful tool for comparing
results known to be generally true with those arising in the quark
model.  This analysis is useful for two reasons: i) It helps to
justify the quark model, which is of utility since the quark model is
the basis of considerable intuition about the excited states; and ii)
it gives direct insight into certain aspects of the phenomenology of
excited states.

The usefulness of the $1/N_c$ expansion for analyzing the properties
of baryons in the ground-state band ({\it i.e.}, the $N$'s and
$\Delta$'s) has been demonstrated for numerous observables including
masses, electromagnetic moments, and axial
couplings~\cite{DM,Jenk,DJM1,DJM2,JL,MR,CGO,list}, with significant
recent interest in properties of the $\Delta$ and the $N$-$\Delta$
transition~\cite{BHL,BL1,BL2,JJM,Coh}.  The spin-flavor symmetry gives
significant insight into the validity of the quark model.  On one hand
it is derived by imposing consistency between the $N_c$ scaling of the
$\pi$-$N$ coupling ($N_c^{1/2}$) and that of $\pi$-$N$ scattering
($N_c^0$)~\cite{GS,DM}.  On the other hand, the simplest way to
compute the consequences of the symmetry is to treat the baryon as
though it were a collection of $N_c$ quark fields in identical
$S$-wave single-particle orbitals and then keep track of the color,
spin, and flavor~\cite{MR,CGO,DJM2}, with the quarks combined into a
color singlet and into a particular representation of combined
spin-flavor symmetry.  The $N_c$ counting can then be implemented by
introducing operators that break the symmetry and are accompanied by a
characteristic power of $1/N_c$.  This organizes operators
contributing to a particular observable according to a well defined
power-counting scheme.

The key point is that for these ground-state band baryons, large $N_c$
QCD has the same spin-flavor symmetry as the simple quark model and
has the same pattern of symmetry breaking.  Thus, at least for those
ground-state baryon properties that are insensitive to the dynamical
details and are essentially fixed by the symmetries, large $N_c$ QCD
provides a justification of the quark model.  In a similar way we wish
to investigate the extent to which large $N_c$ QCD treats excited
baryons in an analogous manner.  That is, we wish to understand what
large $N_c$ rules can be obtained from large $N_c$ consistency rules
at the purely hadronic level, and compare these results to what is
obtained for a quark picture, to see what aspects of the quark picture
can be justified.

A central issue in using the $1/N_c$ expansion to study quantities in
our $N_c\!=\!3$ universe is whether to include quantities that vanish
for $N_c\!=\!3$.  For example, should the quantity $1\!-\!(3/N_c)$ be
evaluated by immediately taking the $N_c \to \infty$ limit and
retained as unity, or should one note that eventually one imposes
$N_c\!=\!3$ in the $1/N_c$ expansion and treat it as vanishing?  This
question was discussed in the case of baryon charge radii and
quadrupole moments in Ref.~\cite{BL2}.  In the present context we
believe it is useful to work in the $N_c \rightarrow \infty$ world
directly, because we are attempting to make {\it qualitative}
comparisons between two different pictures whose large $N_c$ behaviors
are known.  We caution, however, that it is by no means yet entirely
clear how to handle $1/N_c$ corrections to the two pictures in a
consistent way.  This issue is particularly thorny because there exist
states at large $N_c$ that do not exist at large $N_c$, and the role
of these large $N_c$ artifacts must be isolated before one attempts to
draw phenomenological conclusions.

Before starting, we must note a key distinction between excited
baryons and their ground-state band cousins.  The ground-state baryons
are stable in the large $N_c$ limit (Of course, in the real world the
$\Delta$ decays due to the anomalous lightness of the $\pi$, thanks to
approximate chiral symmetry), while excited baryons are all
resonances.  Now, from standard counting based on Witten's original
arguments~\cite{Wit}, it is clear that the characteristic $N_c$
scaling for the excitation energy of an excited baryon is $N_c^0$,
while the scaling of the three-point coupling between the excited
baryon, a ground-state band baryon, and a meson is also $N_c^0$.
Thus, the resonance width also scales as $N_c^0$.  In this respect,
baryons in large $N_c$ QCD are fundamentally different from mesons; in
the meson case we know that widths scale as $1/N_c$, so that
well-defined narrow meson states exist at large $N_c$.  Indeed, from
the perspective of large $N_c$ counting alone, one must be agnostic
about the very existence of baryon resonances that are narrow enough
to isolate.  Here we simply note that the empirical evidence indicates
identifiable resonances.

From the purely hadronic perspective the fact that excited baryons are
associated with resonances simply suggests that the appropriate first
step is to describe scattering processes, such as meson-nucleon
scattering, in channels for which such resonances may reveal
themselves.  The role of large $N_c$ QCD is then simply to relate
scattering in various channels [up to $O(1/N_c)$ corrections], in the
sense that various linear combinations of channels are
equal~\cite{HEHW,MP,MM,Mat3}.  We note that, while these relations
were initially derived in the context of chiral soliton models, they
are in fact model independent.  An outline of a derivation of these
relations directly from large $N_c$ consistency relations along the
lines of Ref.~\cite{DJM2} appears in the Appendix.  From these
relations one can deduce patterns of degeneracy among resonances in
various channels that are valid up to corrections of $O(1/N_c)$.  We
note that, although the linear relations in scattering amplitudes have
been known for a long time, the patterns of mass degeneracies among
the excited baryons reported here are brand new, their existence first
mentioned in our recent Letter~\cite{us}.

The fact that the excited baryons are resonances has always been an
awkward fact for quark models.  If one defines a quark model for a
baryon as a description in which there are $N_c$ constituent quarks
interacting through potentials and with no mechanism for pair
creation, then as a matter of principle there is no way that such a
model can ever describe a physical state that is a meson-nucleon
resonant scattering state.  Implicitly, what is done is to assume that
the quark model describes a state that is relatively weakly coupled
via some quark pair-creation mechanism to a larger Hilbert space that
includes the asymptotic two-hadron state.  If such a coupling is weak
enough, one expects the position of the resonance to be close to the
bound-state energy of the uncoupled system.  Thus, any quark model
treatment that fits parameters so that the energies computed in the
model match resonance masses implicitly makes a weak-coupling
assumption.  This is worth stressing in the present context, if only
to remind ourselves that from a large $N_c$ perspective there is no
reason to assume such a coupling is weak (Note that this is not the
case for mesons, where every type of pair creation is suppressed by
$1/N_c$).  Again, we assume here that the coupling is weak for reasons
not connected to $N_c$, and proceed.

Let us look in a bit more detail at how the quark model for baryons is
realized.  The system can be solved as a true three-body problem, with
the potential consisting of two-body or three-body interactions
between the quarks that depend only on their relative coordinates.
Such a treatment has the virtue of being consistent with the
underlying spirit of the model and has the technical advantage that
the center-of-mass coordinate automatically separates from the
relative coordinates, allowing for a description of internal
excitations that is not contaminated by any spurious center-of-mass
motion.  However, full three-body calculations are technically
difficult.  Moreover, the wave functions are complicated, and thus it
is hard to obtain much intuition from them.  Accordingly, many simple
calculations are based on a single-particle potential-type model,
where this potential is thought of as arising from the interactions of
the other quarks.  In the simplest version of this model (for
$N_c\!=\!3$), the ground-state baryons have all three quarks in the
lowest $S$-wave orbital (yielding a {\bf 56}-plet in spin-flavor for
$N_f\!=\!3$) while the first excited group of baryons has one quark in
a $P$ wave and fills a mixed-symmetry {\bf 70}-plet spin-flavor
state~\cite{close}.  Both types of models can be called quark models,
but for our purposes it is useful to distinguish between them.
Accordingly, we refer to the second variant as the {\it quark-shell
model}, since it has similarities with the shell model of nuclear
physics.

A few additional comments about the quark-shell model are in order.
First, much of the intuition many people have about excited baryons
and much of the language used to describe them are based on the
quark-shell model rather than more sophisticated treatments.  The
reason is that the simplicity of the model allows one to form a
comprehensible picture of the state.  Second, one may create more
sophisticated versions of the quark-shell model that include
admixtures of different single-particle descriptions, in order to
include some of the correlations; such admixtures are called
configuration mixing.  Of course, if all possible configuration mixing
is allowed and if the interactions being used are the full quark-quark
potentials of the underlying quark model, then the quark-shell model
is equivalent to the full three-body quark model, representing just a
convenient basis in which to work rather than a distinct model (The
situation is completely analogous to that of the case in the nuclear
many-body problem~\cite{shell}).  Here, when we refer to the
quark-shell model, we mean models in which configuration mixing is
neglected or taken to be small.  It is also worth noting that the
simple quark-shell model (with little or no configuration mixing) does
a good job of describing the spectrum of the lowest-lying observed
$N$*'s.  In the present context, we note that to date all quark-based
treatments that describe excited baryons in the large $N_c$ limit of
QCD were quark-shell models that neglect configuration
mixing~\cite{CGKM,PY,Goity,CCGL1,CCGL2,GSS,CC1,CC2}.  In such a
picture, the first excited states are either radial excitations of the
symmetric ground-state multiplet (such as the Roper), or orbital
excitations (with quantum number $\ell$) of a single quark with
respect to the other $N_c\!-\!1$ quarks remaining in a spin-flavor
symmetrized ``core.''  Again using SU(6) terminology, these states
fill representations analogous to the three-color {\bf 56} and {\bf
70}, respectively~\cite{CGKM,PY,Goity,CCGL1,CCGL2,GSS,CC1,CC2}.

In this paper we compare the physical content of the two
pictures---excited baryons as resonances in meson-nucleon scattering,
and as single-quark excitations in a quark-shell model---to test
whether the two are consistent with each other in a large $N_c$ world.
We consider the lowest positive- and negative-parity nonstrange
excited baryon resonances.  We find that generically the two pictures
are compatible.  That is, both pictures predict patterns of mass
degeneracy at leading order in the $1/N_c$ expansion and these
patterns are identical.  We note that this compatibility is nontrivial
and may help justify the use of the quark-shell model and explain its
qualitative success.

We should comment briefly on the role of model dependence in what
follows.  The relations that follow from our treatment of meson-baryon
scattering are truly model independent and are direct results of large
$N_c$ QCD.  There is a subtlety in results obtained for excited
baryons using the quark model language: For the ground-state band,
studies of the $N_c$ dependence of operator matrix elements in a quark
model picture~\cite{DJM2} completely reproduce results of the large
$N_c$ consistency conditions.  Something similar can be seen
here---the multiplet structure we obtain using quark model language
below is identical to that obtained by Pirjol and Yan~\cite{PY} using
large $N_c$ consistency relations.  In this sense, their results are
model independent.  However, both the analysis here and that of
Ref.~\cite{PY} are based on treatments of matrix elements of operators
between excited baryon states.  This is strictly only well defined for
stable states.  However, generically, as noted above the excited
baryons are not stable in the large $N_c$ limit; they have widths of
$O(N_c^0)$.

Thus, the relations derived in quark model language are known to be
valid and model independent only for stable excited states.  One could
imagine a world in which the quark masses and the pion were so heavy
that the $N(1535)$ was stable.  In such a world the quark model
results and the large $N_c$ consistency relations would agree for
these stable states and would be truly model independent as $N_c
\rightarrow \infty$.  Unfortunately, with realistic quark masses there
is no reason to think there are any stable baryon resonances at large
$N_c$.  This raises the question of whether any of results are in fact
valid in the real world for the unstable baryons. Of course, it is not
implausible that some results derived in a model-independent way
assuming the states are stable may nevertheless be valid for the
resonances. In essence, the question of whether that is true is at the
heart of this paper.  As we shall see, the multiplet structure of
excited states seen in this quark model-type language, derived
assuming that states are stable, are in fact seen in full large $N_c$
QCD derived without this assumption.  This is one of the principal
results of this paper.

We also use the large $N_c$ results to explore directly aspects of the
phenomenology of the negative-parity baryons, and find two rather
interesting phenomenological results.  The first concerns the
$N_{1/2}$ negative-parity states: The large $N_c$ analysis predicts
the existence of a $N_{1/2}$ negative-parity state whose coupling to
the $\pi$-$N$ channel is weak (vanishing at large $N_c$), but couples
strongly to the $\eta$-N channel.  In fact, $N(1535)$ has precisely
this character.  We similarly predict the other $N_{1/2}$
negative-parity state to couple weakly to the $\eta$-$N$ channel while
coupling strongly to the $\pi$-$N$ channel, which is seen in the
$N(1650)$.  The second result concerns the mixing angles between the
various excited nucleon states in the quark-shell model context.  From
large $N_c$ emerges an analytic result predicting the value of these
mixings, and we find that these predicted values are in good agreement
with phenomenological extractions.

In Sec.~\ref{resonance} we discuss the resonance picture.  The key
point is the existence of model-independent linear relations between
meson-nucleon scattering in various channels that become exact in the
large $N_c$ limit of QCD.  These relations imply degeneracy patterns
for excited baryons.  In Sec.~\ref{quark} we discuss the quark-shell
model picture for the lowest-lying negative-parity baryons.  This
discussion is based on the methods of Refs.~\cite{CCGL1,CCGL2}.
However, we discover an important analytic result not elucidated in
these works, namely, that at leading order in the $1/N_c$ expansion
the mixing angles between various states are fixed and that various
states are degenerate up to this order.  Finally, in
Sec.~\ref{discuss} we discuss the implication of these results both
in justifying the quark-shell model and directly in terms of
phenomenology.

\section{Meson-Nucleon Scattering  Picture} \label{resonance}

It has long been known, primarily through the work of Hayashi, Eckart,
Holzwarth, and Walliser~\cite{HEHW}, and of Mattis and
collaborators~\cite{MP,MM,Mat3,MK}, that the $S$ matrices of various
channels in meson-nucleon scattering (or more generally scattering of
mesons off ground-state band baryons) are linearly related in the
large $N_c$ limit.  For the present purpose it is sufficient to
consider the case of $\pi$ or $\eta$ mesons scattering off a
ground-state band baryon.  In this case the $S$ matrices are given by
\begin{eqnarray}
S_{LL^\prime R R^\prime IJ}^\pi & = &\sum_K (-1)^{R^\prime - R}
\sqrt{(2R+1)(2R^\prime+1)} (2K+1) \left\{ \begin{array}{ccc} K &
I & J\\ R^\prime & L^\prime & 1 \end{array} \right\} \left\{
\begin{array}{ccc} K & I & J \\ R & L & 1 \end{array} \right\}
s_{KL^\prime L}^\pi , \label{MPeqn1} \\
S_{L R J}^\eta & = & \sum_K
\delta_{KL} \, \delta (LRJ) \, s_{K}^\eta .\label{MPeqn2}
\end{eqnarray}
For $\pi$ scattering, the incoming baryon spin (which equals its
isospin) is denoted as $R$, that of the final baryon is denoted
$R^\prime$, the incident (final) $\pi$ is in a partial wave of orbital
angular momentum $L$ ($L^\prime$), and $I$ and $J$ represent the
(conserved) total isospin and angular momentum, respectively, of the
initial and final states (and hence represent isospin and angular
momentum of the state in the $s$ channel).  $S_{LL^\prime R R^\prime
IJ}^\pi$ is the (isospin- and angular momentum-reduced) $S$ matrix for
this channel in the sense of the Wigner-Eckart theorem, the factors in
braces are $6j$ coefficients, and $s_{K L^\prime L}^\pi$ are universal
amplitudes that are independent of $I$, $J$, $R$, and $R'$.  For
$\eta$-meson scattering, since $I_\eta =0$ many of the quantum numbers
are more tightly constrained.  The isospin (= spin) $R$ of the baryon
is unchanged and moreover equals the total isospin $I$ of the state.
The orbital angular momentum $L$ of the $\eta$ remains unchanged in
the process due to large $N_c$ constraints, and $J$ denotes the total
angular momentum of the state, which is constrained by the triangle
rule $\delta (L R J)$.  $S_{L R J}^\eta$ is the reduced scattering
amplitude, and $s_{K}^\eta$ are universal amplitudes independent of
$J$.  The reason that various scattering amplitudes are linearly
related is clear from the structure of Eqs.~(\ref{MPeqn1}) and
(\ref{MPeqn2}): There are more amplitudes $S_{LL^\prime R R^\prime
IJ}^\pi$ than there are $s_{K L^\prime L}^\pi$ amplitudes, and thus
there are linear constraints between them that hold to leading order
in the $1/N_c$ expansion; similarly, there are more $S_{L R J}^\eta$
amplitudes than $s_{K}^\eta$ amplitudes.

Equations~(\ref{MPeqn1}) and (\ref{MPeqn2}) are the starting point for
our analysis of meson-nucleon resonances.  These equations were first
derived in the context of the chiral soliton
model~\cite{HEHW,MP,MK,MM,Mat3}.  In this picture, the quantum number
$K$ has a simple interpretation---the soliton at the classical or
mean-field level breaks both the rotational and isospin symmetries but
preserves the length $K$ of ${\bf K} \equiv {\bf I}+ {\bf J}$.  Thus,
the Hamiltonian describing the intrinsic dynamics of the soliton (that
not associated with collective zero modes) commutes with the ``grand
spin'' ${\bf K}$, and excitations can be labeled by $K$.  It is
important to stress, however, that Eqs.~(\ref{MPeqn1}) and
(\ref{MPeqn2}) are exact results in large $N_c$ QCD and are
independent of any model assumptions.  A derivation directly from the
large $N_c$ consistency rules~\cite{DJM2} exploiting the famous
$I_t=J_t$ rule~\cite{MM} can be found in the Appendix.

The key point for our purposes is that, in order for a resonance to
occur in one channel $S_{LL^\prime R R^\prime IJ}^\pi$ in $\pi$-$N$
scattering, there must be a resonance in at least one of the
contributing amplitudes $s_{K L^\prime L}^\pi$.  However, since the
same $s_{KL^\prime L}^\pi$ contributes to amplitudes in more than one
channel, all of them resonate at the same energy, and this implies
degeneracies in the excited baryon spectrum.  An analogous argument
holds for $\eta$-$N$ scattering.

To make the preceding argument concrete, one needs a method to extract
the resonance position from the $S$ matrix for scattering.  Here we
adopt the usual theoretical prescription: One analytically continues
the scattering amplitude in the complex plane to unphysical but
on-shell kinematics and defines the complex resonance position to be
the position of a pole in the scattering amplitude.  One can then
simply relate the real and imaginary parts to the mass and width of
the resonance.  Using this definition of the resonance position, the
argument given above is quite clean.  Analytically continuing
Eqs.~(\ref{MPeqn1}) and (\ref{MPeqn2}) to the complex plane, it is
apparent that if there is a pole in the complex plane on the left-hand
side for $\pi$-$N$ scattering, then at this pole position the
right-hand side must also diverge, implying a pole in one of the $s_{K
L^\prime L}^\pi$ amplitudes on the right-hand side.  One can then turn
this around and argue, as above, that since $s_{K L^\prime L}^\pi$
contributes to multiple channels, all of them must resonate at the
same point (unless other selection rules, {\it e.g.}, parity, forbid
them to mix); similarly for $\eta$-$N$ scattering.  We note in passing
that, although this theoretical definition of the resonance position
is valid, there is a practical difficulty in extracting the precise
resonance positions from scattering since one cannot directly probe
unphysical kinematics, and thus there is always some model dependence
in any extraction of the resonance position from data.

Consequences of Eqs.~(\ref{MPeqn1}) and (\ref{MPeqn2}) for
negative-parity partial waves can be seen in the right column of
Table~\ref{I}, which lists the linear combinations of the $s^\pi_{K
L^\prime L}$ or $s^\eta_K$ ``$K$-amplitudes'' contributing to a
particular partial wave of fixed $I$ and $J$ in the $s$ channel.  In
using this table to deduce patterns of degeneracy of the baryon
states, it is important to clarify whether any degeneracy might be
expected in the $s_{K L^\prime L}$ and $s^\eta_K$ amplitudes
themselves.  On one hand it is clear that various $K$ sectors ought to
be dynamically distinct, since they are distinct in the large $N_c$
limit.  This is particularly clear in the context of the chiral
soliton models, where the various $K$ sectors are completely separate.
More generally, there is no reason to suspect degeneracy between
different $K$ sectors, and it would be unnatural to impose any such
degeneracies.  On the other hand, it is quite plausible that there may
be degeneracies in the poles of amplitudes with the same $K$ but
different values of $L$ or $L^\prime$: The orbital angular momentum of
the $\pi$ is not an immutable quantity in the same sense as $I$ or
$J$.  Thus, if there is a resonance of fixed $I$ and $J$ but
accessible by various $L$ ({\it e.g.}, by scattering off a $\Delta$
rather than $N$), one would expect resonances in channels of different
$L$ to be degenerate.  Similarly, scattering partial waves involving
different mesons that nevertheless contain amplitudes in the same $K$
channel ({\it e.g.}, $s^\pi_{222}$ and $s^\eta_2$) can produce
degenerate poles.

\begin{table}
\caption{Negative-parity mass eigenvalues in the quark-shell
model picture, corresponding partial waves, and their expansions in
terms of $K$-amplitudes.  The association of masses with states is
from the large $N_c$ quark-shell model relations in Sec.~\ref{quark}.
The superscripts $\pi N N$, $\pi N \Delta$, $\pi \Delta \Delta$, $\eta
N N$, and $\eta \Delta \Delta$ refer to the scattered meson and the
initial and final baryons, respectively.  The partial-wave amplitudes
are derived from Eqs.~(\ref{MPeqn1}) and (\ref{MPeqn2}).  Note that
these states are those appropriate to a large $N_c$ world (As
discussed in the text, some do not occur for $N_c\!=\!3$).  We only
list states with quantum numbers consistent with a single quark
excited to $\ell=1$ and with total isospin of 3/2 or less; and we only
list partial waves sufficient to accommodate all the given resonances
(hence in particular $L = L^\prime \le 2$). \label{I}}
\medskip
\begin{tabular}{lcccccl}
State \mbox{  } && Quark Model Mass \mbox{   } &&
\multicolumn{3}{l}{Partial Wave, $K$-Amplitudes} \\
\hline\hline
$N_{1/2}$ && $m_0$, $m_1$
   && $S^{\pi N N}_{11}$            &=& $s^\pi_{100}$ \\
&& && $D^{\pi \Delta \Delta}_{11}$  &=& $s^\pi_{122}$ \\
&& && $S^{\eta N N}_{1 1}$          &=& $s^\eta_0$ \\
\hline
$\Delta_{1/2}$ && $m_1$, $m_2$
   && $S^{\pi N N}_{31}$            &=& $s^\pi_{100}$ \\
&& && $D_{31}^{\pi \Delta \Delta}$  &=& $\frac{1}{10}
\left( s^\pi_{122} + 9 s^\pi_{222} \right)$ \\
&& && $D^{\eta \Delta \Delta}_{31}$ &=& $s^\eta_2$ \\
\hline
$N_{3/2}$ && $m_1$, $m_2$
   && $D^{\pi N N}_{13}$            &=& $\frac 1 2
\left( s^\pi_{122} + s^\pi_{222} \right)$ \\
&& && $D_{13}^{\pi N \Delta}$       &=& $\frac 1 2
\left( s^\pi_{122} - s^\pi_{222} \right)$ \\
&& && $S_{13}^{\pi \Delta \Delta}$  &=& $s^\pi_{100}$ \\
&& && $D_{13}^{\pi \Delta \Delta}$  &=& $\frac 1 2
\left( s^\pi_{122} + s^\pi_{222} \right)$ \\
&& && $D_{13}^{\eta N N}$           &=& $s^\eta_2$ \\
\hline
$\Delta_{3/2}$  && $m_0$, $m_1$, $m_2$
   && $D^{\pi N N}_{33}$            &=& $\frac{1}{20}
\left( s^\pi_{122} + 5 s^\pi_{222} + 14 s^\pi_{322} \right)$ \\
&& && $D^{\pi N \Delta}_{33}$ &=& $\frac{1}{5\sqrt{10}}
\left( 2 s^\pi_{122} + 5 s^\pi_{222} - 7 s^\pi_{322} \right)$ \\
&& && $S_{33}^{\pi \Delta \Delta}$  &=& $s^\pi_{100}$ \\
&& && $D_{33}^{\pi \Delta \Delta}$  &=& $\frac{1}{25}
\left( 8 s^\pi_{122} + 10 s^\pi_{222} + 7 s^\pi_{322} \right)$ \\
&& && $S_{33}^{\eta \Delta \Delta}$ &=& $s^\eta_0$ \\
&& && $D_{33}^{\eta \Delta \Delta}$ &=& $s^\eta_2$ \\
\hline
$N_{5/2}$ && $m_2$
   && $D^{\pi N N}_{15}$            &=& $\frac{1}{9}
\left( 2 s^\pi_{222} + 7 s^\pi_{322} \right)$ \\
&& && $D_{15}^{\pi N \Delta}$       &=& $\frac{\sqrt{14}}{9}
\left( s^\pi_{222} - s^\pi_{322} \right)$ \\
&& && $D_{15}^{\pi \Delta \Delta}$  &=& $\frac{1}{9}
\left( 7 s^\pi_{222} + 2 s^\pi_{322} \right)$ \\
&& && $D_{15}^{\eta N N}$           &=& $s^\eta_2$ \\
\hline
$\Delta_{5/2}$ && $m_0$, $m_1$
   && $D^{\pi N N}_{35}$            &=& $\frac{1}{90}
\left( 27 s^\pi_{122} + 35 s^\pi_{222} + 28 s^\pi_{322} \right)$ \\
&& && $D_{35}^{\pi N\Delta}$        &=&
$\frac{1}{90} \sqrt{\frac{7}{5}}
\left( 27 s^\pi_{122} + 5 s^\pi_{222} - 32 s^\pi_{322} \right)$ \\
&& && $D_{35}^{\pi \Delta \Delta}$  &=& $\frac{1}{450}
\left( 189 s^\pi_{122} + 5 s^\pi_{222} + 256 s^\pi_{322} \right)$ \\
&& && $D_{35}^{\eta \Delta \Delta}$ &=& $s^\eta_2$ \\
\hline
$\Delta_{7/2}$ && $m_2$
&& $D_{37}^{\pi \Delta \Delta}$ &=& $\frac 1 5
\left( 2 s^\pi_{222} + 3 s^\pi_{322} \right)$ \\
\hline
\end{tabular}
\end{table}

Examples of degenerate negative-parity multiplets at large $N_c$ that
one can infer from the meson-baryon scattering relations include:
\begin{eqnarray}
N_{1/2} , \; \; \;  \Delta_{3/2} , \; \; \; \cdots \; \;  \; \; &~&
 (s_{0}^\eta) , \label{s0}\\
N_{1/2} , \;  \; \;  \Delta_{1/2} , \; \;  \; N_{3/2} , \; \; \;
\Delta_{3/2} , \; \; \;  \Delta_{5/2} , \; \; \; \cdots \; \;  \; \;
 &~& (s_{1 0 0}^\pi, s_{1 2 2}^\pi) , \label{s1}\\
 \Delta_{1/2} , \;  \; \;  N_{3/2} , \;  \; \; \Delta_{3/2} , \;
 \; \; N_{5/2} , \; \; \;  \Delta_{5/2} ,
 \;  \; \; \Delta_{7/2} , \; \; \; \cdots   \; \;
 \; \;  &~& (s_{2 2 2}^\pi, s_{2}^\eta ) , \label{s2} \\
\Delta_{3/2} , \; \; \; N_{5/2} , \; \;  \; \Delta_{5/2} ,
 \; \; \; \Delta_{7/2} , \; \; \ \cdots \; \; \; \; &~&
(s^\pi_{3 2 2} ), \;  \;  \label{s3}
\end{eqnarray}
where the states are listed on the left and the contributing
amplitudes on the right.  The ellipses indicate that in the large
$N_c$ world the multiplets are infinite dimensional, and we have
simply listed the low-spin and -isospin members of the multiplet.  The
preceding multiplets are one of the principle results of this work.  A
few comments about them are in order. First note that at large $N_c$
the poles of the scattering for these various members of the multiplet
occur at the same point in the complex plane.  Thus, the states have
both the same mass and the same width as $N_c \rightarrow \infty$.
Next, let us look at the multiplet in Eq.~(\ref{s1}).  Note that we
have included the $\Delta_{5/2}$ for the $s_{1 0 0}$ channel; this may
be surprising in light of the fact that Table~\ref{I} has no partial
wave for $\Delta_{5/2}$ with a contribution from $s_{1 0 0}$.  It is
because the table is restricted to partial waves in scattering off the
$N$ and $\Delta$, while the $s_{1 0 0}$ contribution to $\Delta_{5/2}$
is seen in scattering off a ground-state band baryon with $I=J=5/2$,
which though absent in our world, exists at large $N_c$.  Note also
that the degeneracy pattern seen in Eq.~(\ref{s1}) occurs for both the
$s_{1 0 0}$ contributions and the $s_{1 2 2}$.  There are two logical
possibilities for this to occur: Either there are two distinct
multiplets with the same degeneracy patterns, or the two multiplets
are in fact the same.  The second possibility is clearly more
economical, and we believe it to be correct.  Similarly, the multiplet
pattern seen in Eq.~(\ref{s2}) occurs for both $\pi$ and $\eta$
scattering, strongly suggesting that the same physical states occur in
both.  These observations tie in neatly with our previous discussion.

We should note that the degeneracy patterns shown above are newly
derived in the context of meson-baryon scattering.  This is somewhat
surprising, since nearly 20 years ago Mattis and Karliner~\cite{MK}
computed the excited baryon spectrum in the Skyrme model directly from
pion-baryon scattering using Eq.~(\ref{MPeqn1}).  In that work the
only degeneracy found was between the $N_{1/2}$ and the
$\Delta_{1/2}$.  How can we understand these degeneracy rules in light
of the fact that explicit calculations based on the same fundamental
formula missed them?  The answer lies in the algorithm used in
Ref.~\cite{MK} to extract resonance positions.  The technique first
computed the scattering amplitudes and then used motion in the Argand
plots to fix the resonance position.  This is essentially the
technique used by experimentalists, and is highly appropriate when
used in comparison with experimental extractions.  However, by
restricting attention to physical kinematics, one cannot directly
access the pole position.

\section{New Results in the Quark-Shell Model Picture}\label{quark}

Although the multiplet structure in Eqs.~(\ref{s0})--(\ref{s3}) is
newly derived here in the context of meson-baryon scattering, the
existence of such towers has been known for some time in the context
of quark model-type treatments in the work or Pirjol and
Yan~\cite{PY}.  Although the formalism used in Ref.~\cite{PY} is
apparently model independent, as discussed in the Introduction it
makes use of a strong dynamical assumption characteristic of the quark
model, namely that the excited states are stable.  This assumption is
generally not compatible with large $N_c$ QCD.  The derivation of
these degenerate multiplets in \cite{PY} is quite beautiful
conceptually, essentially applying the reasoning of
Refs.~\cite{GS,DM}.  However, it is computationally somewhat involved.
Here we rederive these results using a more explicit quark-shell model
language (but an analogous dynamical assumption of stable baryons)
using the methods of~\cite{CCGL1,CCGL2}.

As noted in the Introduction, our approach here is to work directly in
the $N_c \rightarrow \infty$ world.  This observation has a direct
bearing on the enumeration of states in the quark-shell model.  States
in the first negative-parity multiplet ($\ell =1$) have an
($N_c\!-\!1$)-quark core that is completely symmetric under spin
$\times$ flavor, and thus have the quantum numbers $S_c = I_c$ in the
nonstrange case.  Multiple states with the same fixed values of $I$
and $J$ are distinguished by the total spin $S$ carried by the quarks,
and this is denoted~\cite{CCGL1,CCGL2} by primes (no primes for
$S=1/2$, one for $S=3/2$, {\it etc.}).  It is an elementary exercise
in combining angular momenta and isospin to show that, for $N_c \ge
5$, the states with $I \le 3/2$ are $N_{1/2}$, $N^\prime_{1/2}$,
$N_{3/2}$, $N^\prime_{3/2}$, $N^\prime_{5/2}$, $\Delta_{1/2}$,
$\Delta^\prime_{1/2}$, $\Delta_{3/2}$, $\Delta^\prime_{3/2}$,
$\Delta^{\prime\prime}_{3/2}$, $\Delta^\prime_{5/2}$,
$\Delta^{\prime\prime}_{5/2}$, and $\Delta^{\prime\prime}_{7/2}$.  For
$N_c\!=\!3$, the states $\Delta^\prime_{1/2}$, $\Delta^\prime_{3/2}$,
$\Delta^{\prime\prime}_{3/2}$, $\Delta^\prime_{5/2}$,
$\Delta^{\prime\prime}_{5/2}$, and $\Delta^{\prime\prime}_{7/2}$ do
not occur, but must be included in a full $1/N_c$ analysis until the
final step of setting $N_c\!=\!3$.

Working up to $O(N_c^0)$ in the quark-shell model picture, one finds 3
operators contributing to the Hamiltonian, denoted in
Refs.~\cite{CCGL1,CCGL2} as ${\cal H} = c_1 \openone + c_2 \ell s +
c_3 \ell^{(2)} g G_c/N_c$.  To remind the reader (Refer to
Refs.~\cite{CCGL1,CCGL2} if the following notation is not familiar),
lowercase indicates operators acting upon the excited quark, and
subscript $c$ indicates those acting upon the core.  $G^{ia}$ denotes
the combined spin-flavor operator $\propto q^\dagger \sigma^i \tau^a
q$, and $\ell^{(2)}$ is the $\Delta \ell = 2$ tensor operator.  This
is to be contrasted with Refs.~\cite{PY}, which effectively include
only one spin$\times$flavor-breaking operator at this order.

The Hamiltonian up to $O(N_c^0)$ for the mixed $N_{1/2}$ states reads
\begin{equation}
\begin{array}{r} H_{N_{1/2}} = \left(\overline{N}_{1/2} \,
\overline{N}^{\, \prime}_{1/2} \right) {\bf M}_{N_{1/2}} \! \! \! \\
\vphantom{1} \end{array} \left( \begin{array}{c} N_{1/2} \\
N^\prime_{1/2} \end{array} \right) ,
\end{equation}
as may be obtained from Eqs.~(A6)--(A8) or Table~II of
Ref.~\cite{CCGL2}, again including only contributions up to
$O(N_c^0)$.  The mass matrix ${\bf M}_{N_{1/2}}$ is diagonalized by
the unitary matrix $U_{N_{1/2}}$: $U_{N_{1/2}} {\bf M}_{N_{1/2}} \!
U^\dagger_{N_{1/2}} = {\rm diag} \left( M_{N_{1/2}}^{(1)},
M_{N_{1/2}}^{(2)} \right)$, where
\begin{equation} \label{N1mass}
{\bf M}_{N_{1/2}} = \left( \begin{array}{ccc} c_1 N_c -\frac 2 3 c_2 &&
-\frac{1}{3\sqrt{2}} c_2 -\frac{5}{24\sqrt{2}} c_3 \\
-\frac{1}{3\sqrt{2}} c_2 -\frac{5}{24\sqrt{2}} c_3 && c_1 N_c -\frac 5
6 c_2 -\frac{5}{48} c_3 \end{array} \right),
\end{equation}
\begin{equation}
U_{N_{1/2}} = \left( \begin{array}{rr} \cos \theta_{N_1} &
\sin \theta_{N_1} \\ -\sin \theta_{N_1} & \cos \theta_{N_1}
\end{array} \right) ;
\end{equation}
$N_{1/2}$ and $N^\prime_{1/2}$ refer to unmixed negative-parity
spin-1/2 nucleon states in the initial quark-shell model basis.
Anticipating a remarkable result, let us define 3 particular
combinations of the parameters:
\begin{eqnarray}
m_0 & \equiv & c_1 N_c - \left( c_2 + \frac{5}{24} c_3 \right) ,
\nonumber \\
m_1 & \equiv & c_1 N_c - \frac 1 2 \left( c_2 - \frac{5}{24} c_3
\right), \nonumber \\
 m_2 & \equiv & c_1 N_c + \frac 1 2 \left( c_2 -
\frac{1}{24} c_3 \right) .
\end{eqnarray}
One finds that $M_{N_{1/2}}^{(1)} = m_0$, $M_{N_{1/2}}^{(2)} = m_1$,
and $\tan \theta_{N_1} = \sqrt{2}$.  Note first that, had the
numerical coefficients in Eq.~(\ref{N1mass}) been arbitrary, the
eigenvalues would in general contain square roots of terms quadratic
in $c_2$ and $c_3$, and the mixing angle would have been a complicated
function of their ratio.  The simplicity of the actual results
indicates something deep is happening. Indeed, this simple mixing
angle result at large $N_c$ was earlier noticed in the work of Pirjol
and Yan~\cite{PY}. As will be seen shortly, results for states with
other quantum numbers are equally simple.

Note that the simplicity of the present result---analytic expressions
for both the masses and the mixing angle---was not noted in
Refs.~\cite{CCGL1,CCGL2}; the reason is simply that previous work
always included the $O(1/N_c)$ terms in the Hamiltonian and then
diagonalized numerically.  The simple result given above, however,
only holds to $O(N_c^0)$, and the inclusion of the $1/N_c$ correction
terms in the Hamiltonian obscured the simple leading result.

Using an analogous notation for the other states,
\begin{equation}
{\bf M}_{N_{3/2}} = \left( \begin{array}{ccc} c_1 N_c +\frac 1 3 c_2 &&
-\frac{\sqrt{5}}{6} c_2 +\frac{\sqrt{5}}{48} c_3 \\
-\frac{\sqrt{5}}{6} c_2 +\frac{\sqrt{5}}{48} c_3 && c_1 N_c -\frac 1 3
c_2 +\frac{1}{12} c_3 \end{array} \right) ,
\end{equation}
we find $M_{N_{3/2}}^{(1)} = m_2$, $M_{N_{3/2}}^{(2)} = m_1$ and $\tan
\theta_{N_3} = -1/\sqrt{5}$.  The spin-5/2 state is unmixed but also
has a degenerate eigenvalue: $M_{N_{5/2}} = m_2$.

Next let us consider the $\Delta$ states:
\begin{equation}
{\bf M}_{\Delta_{1/2}} = \left( \begin{array}{ccc} c_1 N_c +\frac 1 3
c_2 && +\frac{\sqrt{5}}{6} c_2 -\frac{\sqrt{5}}{48} c_3 \\
+\frac{\sqrt{5}}{6} c_2 -\frac{\sqrt{5}}{48} c_3 && c_1 N_c -\frac 1
3 c_2 +\frac{1}{12} c_3 \end{array} \right) ,
\end{equation}
which gives $M_{\Delta_{1/2}}^{(1)} = m_2$, $M_{\Delta_{3/2}}^{(2)} =
m_1$, and $\tan \theta_{\Delta_1} = -1/\sqrt{5}$.  Next,
\begin{equation}
{\bf M}_{\Delta_{3/2}} = \left( \begin{array}{ccccc} c_1 N_c -\frac 1 6
c_2 && +\frac{5}{6\sqrt{2}} c_2 + \frac{1}{48\sqrt{2}} c_3 &&
+\frac{\sqrt{3}}{16} c_3 \\ +\frac{5}{6\sqrt{2}} c_2 +
\frac{1}{48\sqrt{2}} c_3 && c_1 N_c - \frac{2}{15} c_2 - \frac{1}{15}
c_3 && -\frac{3}{10} \sqrt{\frac{3}{2}} c_2 - \frac{7}{80}
\sqrt{\frac{3}{2}} c_3 \\ +\frac{\sqrt{3}}{16} c_3 && -\frac{3}{10}
\sqrt{\frac{3}{2}} c_2 - \frac{7}{80} \sqrt{\frac{3}{2}} c_3 && c_1 N_c
-\frac{7}{10} c_2 - \frac{7}{120} c_3 \end{array} \right) ,
\end{equation}
with $M_{\Delta_{3/2}}^{(1)} = m_0$, $M_{\Delta_{3/2}}^{(2)} = m_2$,
and $M_{\Delta_{3/2}}^{(3)} = m_1$.  The mixing angles are
parameterized by
\begin{equation}
U_{\Delta_{3/2}} = \left( \begin{array}{ccccc}
\cos\theta \cos\phi && \sin\theta \cos\phi && \sin\phi \\
-\cos\theta \sin\phi \sin\psi -\sin\theta \cos\psi && -\sin\theta
\sin\phi \sin\psi + \cos\theta \cos\psi && \cos\phi \sin\psi \\
-\cos\theta \sin\phi \cos\psi +\sin\theta \sin\psi && -\sin\theta
\sin\phi \cos\psi -\cos\theta \sin\psi && \cos\phi \cos\psi
\end{array} \right) ,
\end{equation}
and turn out to be $\tan \theta = -\sqrt{2}$, $\tan \phi = -1$, and
$\tan \psi = -1/3$.  $\Delta_{5/2}$ has
\begin{equation}
{\bf M}_{\Delta_{5/2}} = \left( \begin{array}{cccc} c_1 N_c + \frac 1 5
c_2 + \frac{1}{60} c_3 && -\frac{\sqrt{21}}{10} c_2 +
\frac{\sqrt{21}}{80} c_3 \\ -\frac{\sqrt{21}}{10} c_2 +
\frac{\sqrt{21}}{80} c_3 && c_1 N_c -\frac 1 5 c_2 + \frac{1}{15} c_3
\end{array} \right) ,
\end{equation}
from which $M_{\Delta_{5/2}}^{(1)} = m_0$, $M_{\Delta_{5/2}}^{(2)} =
m_1$, and $\tan\theta_{\Delta_5} = \sqrt{3/7}$.  Finally,
$\Delta_{7/2}$ is unmixed and has eigenvalue $M_{7/2} = m_2$.  The
pattern, masses being equal to one of three eigenvalues and mixing
angles having simple expressions, obviously extends into the $\Delta$
sector.  The compilation of mass eigenvalues into states of various
quantum numbers is given in Table~\ref{I}.

Clearly, the fact that all of the masses described by the model are
given by either $m_0$, $m_1$, or $m_2$ to leading order in the $1/N_c$
expansion implies that at large $N_c$ the various states fall into
degenerate multiplets.  These multiplets are given by
\begin{eqnarray}
N_{1/2} , \; \; \;  \Delta_{3/2} , \; \; \; \cdots \; \;  \; \; &~&
 (m_0) , \label{m0}\\
N_{1/2} , \;  \; \;  \Delta_{1/2} , \; \;  \; N_{3/2} , \; \; \;
\Delta_{3/2} , \; \; \;  \Delta_{5/2} , \; \; \; \cdots \; \;  \; \;
 &~& (m_1) , \label{m1}\\
 \Delta_{1/2} , \;  \; \;  N_{3/2} , \;  \; \; \Delta_{3/2}, \;
 \; \; N_{5/2} , \; \; \;  \Delta_{5/2} ,
 \;  \; \; \Delta_{7/2} , \; \; \; \cdots   \; \;
 \; \;  &~& (m_2 ) , \label{m2}
\end{eqnarray}
where the states are listed on the left and the masses on the right.

After a draft of the present paper was completed we became aware of
similar work by Pirjol and Schat~\cite{PS}, who obtained similar
results and computed $1/N_c$ corrections.

\section{Discussion} \label{discuss}

\subsection{On the Compatibility of the Resonance and Quark-Based
Pictures}

The results of the previous two sections are quite striking.  In both
the resonance pole picture and the quark-shell model picture, one
finds that the excited baryons are organized into multiplets of states
that are degenerate modulo splittings arising at next-to-leading
order, $O(1/N_c)$.  Let us compare the multiplet structures predicted
by the two pictures as given in Eqs.~(\ref{s0})-(\ref{s3}) and
Eqs.~(\ref{m0})-(\ref{m2}).  It is clear that the multiplet structure
in Eq.~(\ref{s0}) is identical to that of Eq.~(\ref{m0}); that in
Eq.~(\ref{s1}) is identical to that of Eq.~(\ref{m1}); and that in
Eq.~(\ref{s2}) is identical to that of Eq.~(\ref{m2}).  The two
pictures are compatible.  The interpretation is quite simple: the
multiplet states with mass $m_0$ in the quark-shell model are those
states for which the resonance occurs in the $K=0$ scattering channel
and analogously for $m_1$ states ($K=1$) and $m_2$ states ($K=2$).
This result is highly significant in justifying the quark model.

Before turning to the question of just how strong this justification
is, we note that the resonance picture can have poles for $K=3$, as
seen in Eq.~(\ref{s3}), for which we have not reported any analogous
states in the quark-shell model.  This in no way spoils the
compatibility of the two pictures.  Rather, it merely reflects the
fact that the quark-shell model studies were limited to single-quark
excitations in the lowest $\ell=1$ orbital.  Had we considered higher
excitations in a quark-shell model, such as two-quark excitations or
excitations in the $\ell=3$ orbital, presumably we would have seen a
degenerate multiplet consistent with Eq.~(\ref{s3}).  This prediction,
that such a multiplet structure will be found higher in the spectrum
of a large $N_c$ quark-shell model, is a stringent test of our
interpretation.  Studies of higher excited states in the quark-shell
model are presently under way~\cite{future}.

While studies of higher negative-parity states remain to be completed,
we have also computed the degeneracy patterns for the lowest-lying
excited positive-parity states.  These states include $N(1440)$
$(P_{11})$ and $\Delta(1600)$ ($P_{33}$).  One finds, using
Eq.~(\ref{MPeqn1}), the relations
\begin{eqnarray}
P_{11}^{\pi NN} & = & (s^\pi_{011} + 2s^\pi_{111})/3 , \nonumber \\
P_{13}^{\pi NN} & = & (s^\pi_{111} + 5s^\pi_{211})/6 , \nonumber \\
P_{31}^{\pi NN} & = & (s^\pi_{111} + 5s^\pi_{211})/6 , \nonumber \\
P_{33}^{\pi NN} & = & (2s^\pi_{011} + 5s^\pi_{111} + 5s^\pi_{211})/12
.
\end{eqnarray}
Since no poles are observed in the lowest multiplet for $P_{13}$ or
$P_{31}$, consistency between the resonance and quark-shell model
pictures is achieved simply by placing a pole in $s^\pi_{011}$ but not
in $s^\pi_{111}$ or $s^\pi_{211}$ for this multiplet.  We note that
this result holds more generally, being valid for any state that in
the quark-shell model lies in a spin-flavor symmetric multiplet.

The present interpretation---that quark shell model states correspond
to a well-defined $K$ quantum number (up to $1/N_c$ corrections)---is
based on meson-baryon scattering, with scattering restricted to
$\pi$-baryon and $\eta$-baryon channels.  Of course, if the
interpretation is correct, one should also find consistency between
the multiplet structure of the quark-shell model and resonances
deduced from mixed scattering, in which the initial meson is a $\pi$
and final meson is an $\eta$.  Although not presented here, it is
straightforward to verify that this is true.

Let us now turn to the question of just how strongly the present
result justifies the quark-shell model picture.  We start by observing
that, whatever justification there is for the quark-shell model at
$N_c\!=\!3$ should become increasingly reliable as $N_c$ becomes
large.  The basic point is simply that there are more quarks available
that combine to generate the effective single-body potential seen by
the last quark.  Thus, if there is justification for the model at
$N_c\!=\!3$, it is likely stronger for large $N_c$.  Conversely, a
clear failure of the model at large $N_c$ would suggest that the
picture is unlikely to be valid for $N_c\!=\!3$.  Thus, the fact that
the quark-shell picture produces the same qualitative spectrum in
terms of the multiplet structure as is seen in large $N_c$ QCD from
scattering is a real test, in the sense that the failure to do so
would have cast serious doubts on the model.

Of course, the fact that the multiplet structure in the quark-shell
model agrees with that of large $N_c$ QCD does not justify all aspects
of the model.  In particular, it does not justify the dynamical
details of the model in general, and certainly does not justify an
approach that neglects open channels for decay.  Rather, as for the
ground-state band, it justifies those aspects of the model that
essentially follow from the contracted SU($2N_f$) symmetry.

Finally, we note that compatibility of the excitation spectrum of the
quark-shell model with large $N_c$ QCD is highly nontrivial.  At first
blush, one might think the result {\em is\/} trivial.  After all,
Witten's initial derivation~\cite{Wit} of large $N_c$ rules for
baryons was done using heavy quarks, which essentially defines a quark
model in the first sense described in the Introduction.  Moreover, the
counting used by Witten was essentially combinatoric and applied
independently of the dynamical details.  Thus, one expects that any
relations that apply for large $N_c$ QCD should hold in generic large
$N_c$ quark models.

However, as noted in the Introduction, we are {\em not\/} studying a
full quark model; rather we are using a quark-shell model.  Of course,
one can again appeal to Witten's original argument and argue that the
Hartree approximation emerging at large $N_c$ is a single-particle
picture in exactly the same manner as the quark-shell model.  However,
the preceding argument is specious.  It is certainly true that the
Hartree approximation becomes valid at large $N_c$ {\em for the ground
state}.  Whether it is valid for excited states is a bit more subtle.
It is well known~\cite{many} in many-body theory that mean-field
theories such as the Hartree approach respect the underlying
symmetries of the theory or spontaneously break them.  However, the
use of a mean-field potential for excited states, for example as in
the Tamm-Dancoff approximation, involves {\it ad hoc\/} truncations
that generically violate the symmetries.  In contrast,
symmetry-conserving approximations to treat excited states, such as
linear response theory or random phase approximation, typically go
beyond the mean-field potential and take into account ground-state
correlations.  Since the results of our studies are essentially group
theoretic---and thus entirely dependent on the treatment of the
symmetries---it is by no means {\it a priori} obvious that the use of
the mean-field potential for treatments of the excited states is
adequate.  Thus, the success of the quark-shell model in replicating
the multiplet structure is quite significant.  Again we note that it
is important that the quark-shell model is justified (and not just the
full treatment of the $N_c$-body quark model), as much intuition has
been gleaned from the quark-shell model over the years.

\subsection{Phenomenological Consequences}

The results of Secs.~\ref{resonance} and \ref{quark} may be used to
gain phenomenological insight into the low-lying excited baryons.  A
certain amount of care must be exercised when doing this, however,
since the physical world of $N_c\!=\!3$ cannot be regarded as an
approximately large $N_c$ world for all purposes.  In particular,
consider the most striking formal results of this work---the existence
of nearly degenerate multiplets of states associated with a fixed $K$.
Unfortunately, it will be very difficult to extract this structure
directly in the baryon spectrum for the low-lying odd-parity states in
the real world.  The key difficulty concerns the scales of the
problem: Note that the physical states in question vary in mass from
1520 MeV ($N_{3/2}$) to 1700 ($\Delta_{3/2}$), as listed by the
Particle Data Group~\cite{PDG}.  Thus, all of the observed states lie
within a 200 MeV window.  However, $\Delta$-$N$ mass splitting, a
$1/N_c$ effect, is $\sim 290$~MeV.  Thus, the actual splittings are
not large enough to resolve, given the characteristic size of the
$1/N_c$ effects.  Studies of constraints on these next-to-leading
$1/N_c$ effects are underway~\cite{future}.

Fortunately, there are phenomenological predictions of the preceding
analysis that may well be meaningful in the physical world.  Consider
Eqs.~(\ref{s0}) and (\ref{m0}).  With the interpretation above, we
would say that the quark-shell states with mass $m_0$ in the large
$N_c$ world correspond to states accessible in scattering experiments
characterized by modes with $K=0$.  Now, it so happens that $\pi$-$N$
scattering does not couple to negative-parity $K=0$ modes.  This can
be seen directly from the structure of the $6j$ coefficients in
Eq.~(\ref{MPeqn1}), which implies that $K \ge |L -1|$.  Clearly, $K=0$
can only happen for $L=1$, but $L=1$ makes even-parity states.  Thus,
in Eq.~(\ref{s0}) we see that the $K=0$ multiplet is accessible via
$\eta$-N scattering but not via $\pi$-$N$.  Of course, this result
that the $K=0$ negative-parity states couple to the $\eta$-N channel
but not to the $\pi$-$N$ only holds to leading in order in $1/N_c$, so
the actual prediction is that the coupling to the $\pi$-$N$ channel is
weak for these states.  In Eq.~(\ref{s0}) we list two such states in
the multiplet, $N_{1/2}$ and $\Delta_{3/2}$.  Of these, only the
$N_{1/2}$ can be clearly discerned at $N_c\!=\!3$.  Recall that at
large $N_c$ there are three $\Delta_{3/2}$ negative-parity states in
the quark-shell model with a single $\ell=1$ quark excitation, but for
$N_c\!=\!3$ there is only one.  Thus, one cannot associate the
physical negative-parity $\Delta_{3/2}$ state with a given $K$.  In
contrast, there are two negative-parity $N_{1/2}$ states, both at
large $N_c$ and for $N_c\!=\!3$.  Thus we predict that, to the extent
the $1/N_c$ expansion is useful, one of these two states couples
weakly to pions.  Similarly, the other state has $K=1$, and by an
analogous argument, it should be clear that this state couples
strongly to the $\pi$-$N$ channel but weakly to the $\eta$-$N$
channel.

How well are these prediction borne out in nature?  According to the
Particle Data Group~\cite{PDG}, the $N(1535)$ has a decay fraction to
the $\pi$-$N$ channel (35--55\%) that is virtually equal to its decay
fraction to the $\eta$-$N$ channel (30--55\%).  Now, this is striking
since the phase space factor for the decay is nominally $\sim 2.6$
times larger for the $\pi$-$N$ channel.  Moreover, the nominal phase
space obtained by assuming that the decaying particle has its
Breit-Wigner mass presumably understates the relative advantage the
pion has in phase space: Since the $\eta$-$N$ threshold is only 50 MeV
from the nominal mass of the $N(1535)$, if one averages over the width
of the resonance in estimating the effective phase space, one
substantially reduces the phase space for the $\eta$-$N$ channel but
not for the $\pi$-$N$ channel.  Thus, the $N(1535)$ clearly is much
more strongly coupled to the $\eta$-$N$ channel than to $\pi$-$N$.
Next, consider the $N(1650)$; according to the Particle Data
Group~\cite{PDG}, $N(1650)$ has a decay fraction to the $\pi$-$N$
channel (55--90\%) that is much larger than the decay fraction to the
$\eta$-$N$ channel (3--10\%).  The dominance of the $\pi$-$N$ channel
is far larger than what one estimates purely from the phase space,
since the $\pi$-$N$ channel has a phase space factor $\sim 1.6$ times
that of the $\eta$-$N$ channel.  Thus, the qualitative large $N_c$
predictions for the dominant decay modes of the $N_{1/2}$ states are
apparent in the data.

The success in describing the decay modes of the $N_{1/2}$ states tell
us that the leading large $N_c$ result does a good job in describing
the mixing between these two states.  One can also ask about the
mixing in the context of the quark-shell model, where it is
parameterized by mixing angles.  Thus, another check on the
phenomenological usefulness of the leading-order large $N_c$ treatment
is its ability to predict these mixing angles in a manner that is
qualitatively consistent with traditional quark models.  Again, we
restrict our attention to those sectors for which large $N_c$ and
$N_c\!=\!3$ have the same number of quark model states---namely the
$N_{1/2}$ and $N_{3/2}$ states.  From Sec.~\ref{quark} we see that the
mixing angles are given by $\theta_{N_1} = \tan^{-1}(\sqrt{2})\approx
0.96$ and $\theta_{N_3} = \tan^{-1}(1/\sqrt{5})\approx 2.72$ (angles
in radians).  In comparison, fits using decay data give $\theta_{N_1}
= 0.56$, $\theta_{N_3} = 2.96$ using SU(6)~\cite{HLC}.  The fact that
$N_1$ mixing in large $N_c$ is not too far from the phenomenological
one is, of course, not surprising since the latter was fit to decays;
as we have seen, this is qualitatively consistent with the large $N_c$
result for the $N_{1/2}$ states.  It is encouraging, however, to note
that the phenomenological fits for the $N_{3/2}$ also work well.  We
note also that similar results were obtained directly from $N_c\!=\!3$
quark model calculations~\cite{IK}, as well as large $N_c$ quark-shell
model approaches in which $1/N_c$ corrections are included within
individual operator matrix elements~\cite{CGKM,CCGL1,CCGL2,CC1,CC2}.
It will be interesting to see if similar predictions can be obtained
in the strange sector~\cite{future}.
 
\acknowledgments
The authors acknowledge valuable discussions with Dan Pirjol.  T.D.C.\
acknowledges the support of the U.S.~Department of Energy through
grant DE-FG02-93ER-40762.  R.F.L.\ thanks William Kaufmann for a
useful $6j$ and $9j$ program, and acknowledges support from the
National Science Foundation under Grant No.\ PHY-0140362.

\appendix
\section{Model-independent Large $N_c$ Relations For Meson-Baryon
Scattering}

As mentioned in Sec.~\ref{resonance}, the relations between various
channels in pion-baryon scattering in Eq.~(\ref{MPeqn1}) were first
derived in the context of a chiral soliton
model~\cite{HEHW,MP,MK,MM,Mat3}.  However, the result is exact in the
large $N_c$ limit of QCD.  In this Appendix we show how this result
can be derived directly from large $N_c$ consistency rules with
essentially no model-dependent assumptions.  Before doing this, we
note that it has long been recognized these results are in fact
model independent in the large $N_c$ limit.

As noted long ago by Adkins and Nappi~\cite{AN}, the Skyrme model
(treated via semi-classical projection, as appropriate for the large
$N_c$ limit of QCD) has the striking feature that many relations are
completely independent of the details of the Skyrme model
Lagrangian---{\it i.e.}, they are independent of the values of the
parameters in the Lagrangian, the number and type of terms included in
the Lagrangian, or even the number and types of degrees of freedom.
For example, the $\pi$-$N$ coupling is precisely 2/3 of the
$\pi$-$N$-$\Delta$ coupling in the leading-order treatment of {\it
any} chiral soliton model.  It was a reasonable conjecture that such
relations are fully model independent and are exact results of QCD in
the large $N_c$ limit.

This conjecture was subsequently shown to be correct for all such
relations tested.  The method used the large $N_c$ consistency rules
discovered by Gervais and Sakita in the 1980's~\cite{GS} and then
rediscovered and greatly extended by Dashen and Manohar~\cite{DM} in
the 1990's.  The only assumptions used with this method are: i)
Baryonic quantities in large $N_c$ QCD scale according to the generic
large $N_c$ rules of Witten~\cite{Wit}, or more slowly (if there are
cancellations); ii) there exists a hadronic description that
reproduces the large $N_c$ QCD results; iii) the $\pi$-$N$ coupling
scaling is generic (without cancellations), scaling as $N_c^{1/2}$;
and iv) nature is realized in the most symmetric representation of the
contracted SU(4) group that emerges from the previous assumptions.
The key to the method is to compare the scaling of $\pi$-$N$
scattering (which scales as $N_c^0$) with the $\pi$-$N$ coupling
constant (which scales as $N_c^{1/2}$), suggesting that the sum of the
Born term in $\pi$-$N$ scattering plus the cross graph scales as
$N_c^1$.  This mismatch in scattering implies the need for
cancellations, and these in turn are only possible if the baryons form
nearly degenerate bands of states with $I=J$ (in the two-flavor case)
and lie in irreducible representations of a contracted SU($2N_f$).
The consequences of this symmetry for various matrix elements are
worked out in detail in a series of papers by Dashen, Jenkins, and
Manohar~\cite{DJM1}.  One consequence is that all of the relations
derived in the Skyrme model, but that are insensitive to Skyrme model
details, are in fact results of large $N_c$ QCD.

Now we note that that Eqs.~(\ref{MPeqn1}) and (\ref{MPeqn2}) are, in
fact, completely independent of the details of the Skyrme model, and
thus one expects that it should be possible to derive it using the
methods of Ref.~\cite{DJM1}.  We focus first on the case of $\pi$
scattering, hence the explicit $I_\pi = 1$ below.  The first step is
to express the amplitudes in terms of $t$-channel rather than
$s$-channel exchange.  This can be done simply via a standard relation
between $6j$ coefficients~\cite{edmonds}:
\begin{eqnarray}
 &~&\left\{  \begin{array}{ccc} K & I & J
\\ R^\prime & L^\prime & 1 \end{array} \right\} \left\{
\begin{array}{ccc} K & I & J \\ R & L & 1 \end{array}\right \}  =
\nonumber \\ &~ &\sum_{\cal J} \, (-1)^{I + J + L + L^{\prime} + R +
R^{\prime} + K + {\cal J}} \, (2 {\cal J} +1) \, \left\{
\begin{array}{ccc} 1 & R^\prime & I
\\ R & 1 & {\cal J} \end{array} \right\}
\left\{ \begin{array}{ccc} R^\prime & J &
L^\prime \\ L & {\cal J} & R \end{array} \right\} \left\{
\begin{array}{ccc} 1 & L^\prime & K
\\ L & 1 & {\cal J} \end{array} \right\} .
\label{ident}
\end{eqnarray}
Inserting this identity into the first of Eqs.~(\ref{MPeqn1}) yields
\begin{eqnarray}
&~&S_{L L^\prime R R^\prime I J}  = \sum_{\cal J} \,\left\{
\begin{array}{ccc} 1 & R^\prime & I
\\ R & 1 & {\cal J} \end{array} \right\} \left\{ \begin{array}{ccc}
R^\prime & J & L^\prime \\ L & {\cal J} & R \end{array} \right\}
s_{{\cal J} L L^\prime}^t ,
\label{IequalsJ1}
\end{eqnarray}
with
\begin{eqnarray}
&{}& s_{{\cal J} L L^\prime }^t \equiv \sum_K \, (-1)^{J + I + L +
L^\prime +K + {\cal J} +1} (2 {\cal J} +1) (2K+1)\sqrt{(2R+1)
(2R^\prime +1 )} \,
\left\{\begin{array}{ccc} 1 & L^\prime & K
\\ L & 1 & {\cal J} \end{array} \right\} s^\pi_{K L L^\prime} .
\label{IequalsJ2}
\end{eqnarray}
From the first $6j$ coefficient in Eq.~(\ref{IequalsJ1}), it is
apparent that ${\cal J}$ is the isospin exchanged in the $t$-channel,
while the second $6j$ coefficient implies that ${\cal J}$ is angular
momentum exchanged in the $t$ channel.  The superscript $t$ in the
function $s_{{\cal J} L L^\prime }^t$ indicates that this function is
given in terms of the angular momentum and isospin in the $t$-channel.
The fact that $t$ channel-exchanged isospin and the $t$
channel-exchanged angular momentum are both equal to ${\cal J}$, of
course, implies that they are equal to each other.  Thus
Eq.~(\ref{IequalsJ1}) encodes the celebrated $I_t=J_t$ rule of Mattis
and Mukerjee~\cite{MM}.

Now, the preceding argument shows that Eq.~(\ref{MPeqn1}) implies the
$I_t=J_t$ rule for pion-baryon scattering.  For our purpose, the
important point is that the converse is also true: The $I_t=J_t$ rule
implies Eq.~(\ref{MPeqn1}), since the right-hand side of
Eq.~(\ref{IequalsJ1}) is the most general form for a scattering
amplitude consistent with $I_t = J_t$.  Thus, if one can establish the
$I_t=J_t$ rule directly from the large $N_c$ consistency rules, then
one has established Eq.~(\ref{MPeqn1}) directly from large $N_c$ QCD.
However, as shown in Ref.~\cite{KapMan} it is straightforward to
establish the $I_t=J_t$ rule using the techniques of Ref.~\cite{DJM1}.
An analogous argument works for the case of $\eta$-baryon scattering.

\end{document}